\documentclass[DIV=calc,paper=a4,fontsize=11pt,twocolumn]{scrartcl} 

\usepackage[english]{babel}
\usepackage[protrusion=true,expansion=true]{microtype}
\usepackage{amsmath,amsfonts,amsthm}
\usepackage[final]{graphicx}
\usepackage{xcolor}
\usepackage[normal,small,hypcap,up,labelfont=bf,textfont=it]{caption}
\usepackage{epstopdf}
\usepackage{subfig}
\usepackage{booktabs}
\usepackage{fix-cm}
\usepackage{amssymb,amsfonts}
\usepackage{dsfont}
\usepackage{bbm}
\usepackage{pstricks}
\usepackage{cite}
\usepackage[utf8]{inputenc}
\usepackage[perpage,symbol*]{footmisc}
\usepackage[varg]{txfonts}
\usepackage{balance}
\usepackage{fancyhdr}
\PassOptionsToPackage{hyphens}{url}\usepackage[pdfencoding=auto,psdextra]{hyperref}
\usepackage{bookmark}
\usepackage{verbatim}
\usepackage{fontenc}
\usepackage{cuted}
\usepackage{widetext}

\usepackage{bm}
\usepackage{mathrsfs}

\theoremstyle{definition}

\theoremstyle{plain}

\DeclareCaptionFont{mycolor}{\color[HTML]{000000}}
\usepackage{sectsty}													
\allsectionsfont{
\color[HTML]{31ADF3}\usefont{OT1}{phv}{b}{n}
}

\sectionfont{
\color[HTML]{31ADF3}\usefont{OT1}{phv}{b}{n}
}

\usepackage{fancyhdr}												
\usepackage{lettrine}
\newcommand{\initial}[1]{%
\lettrine[lines=3,lhang=0.3,nindent=0em]{
\color[HTML]{31ADF3}
{\textsf{#1}}}{}}

\usepackage{titling}															

\newcommand{\HorRule}{\color[HTML]{31ADF3}
\rule{\linewidth}{1pt}%
}

\pretitle{\vspace{-30pt} \begin{flushleft} \HorRule
\fontsize{34}{34} \usefont{OT1}{phv}{b}{n} \color[HTML]{31ADF3} \selectfont
}
\title{Coherence, Interference and Visibility}					
\posttitle{\par\end{flushleft}\vskip 0.5em}

\preauthor{\begin{flushleft}\large \lineskip 0.5em \usefont{OT1}{phv}{b}{sl} \color[HTML]{31ADF3}}
\author{Tabish Qureshi\\[8pt]}											
\postauthor{\footnotesize \usefont{OT1}{phv}{m}{sl} \color[HTML]{000000}
Centre for Theoretical Physics, Jamia Millia Islamia, New Delhi,
India. E-mail: \href{mailto:tabish@ctp-jamia.res.in}{tabish@ctp-jamia.res.in}\\[10pt]		
\scriptsize\usefont{OT1}{phv}{m}{n} \color[HTML]{31ADF3}{\textbf{Editors: \emph{Louis Marchildon} \& \emph{Danko Georgiev}} }\\[5pt]
\color[HTML]{000000}{Article history: Submitted on May 10, 2019;  Accepted on June 13, 2019; Published on June 17, 2019.}
\par\end{flushleft}\HorRule}

\date{}																				

\begin{document}
\maketitle
\thispagestyle{fancy} 			
\initial{T}\textbf{he interference observed for a quanton, traversing more than one path, is believed to characterize its wave nature. Conventionally, the sharpness of interference has been quantified by its visibility or contrast, as defined in optics. Based on this visibility, wave-particle duality relations have been formulated for two-path interference. However, as one generalizes the situation to multi-path interference, it is found that conventional interference visibility is not a good quantifier. A recently introduced measure of quantum coherence has been shown to be a good quantifier of the wave nature. The subject of quantum coherence, in relation to the wave nature of quantons and to interference visibility, is reviewed here. It is argued that coherence can be construed as a more general form of interference visibility, if the visibility is measured in a different manner, and not as contrast.\\ Quanta 2019; 8: 24--35.}

\begin{figure}[b!]
\rule{245 pt}{0.5 pt}\\[3pt]
\raisebox{-0.2\height}{\includegraphics[width=5mm]{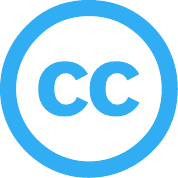}}\raisebox{-0.2\height}{\includegraphics[width=5mm]{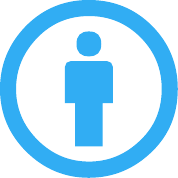}}
\footnotesize{This is an open access article distributed under the terms of the Creative Commons Attribution License \href{http://creativecommons.org/licenses/by/3.0/}{CC-BY-3.0}, which permits unrestricted use, distribution, and reproduction in any medium, provided the original author and source are credited.}
\end{figure}

\section{Introduction}

The phenomenon of interference was discovered long back in 1801 when
Thomas Young performed his seminal experiment by making a light beam
pass through two spatially separated paths, and observing bright and
dark bands on the screen signifying interference \cite{young}. This was
a follow-up of the wave theory of light which he had been developing.
Interference was understood as resulting from superposition of two waves,
which add constructively or destructively at different locations. It was
soon realized that in order to produce an observable pattern of interference
fringes, the two waves must have a constant phase difference between them.
This property of having a constant phase difference between two
waves was called {\em coherence}. If the phase difference between two 
waves is not constant, one has to specify how much does the phase difference
vary with time, or the degree of coherence. The degree of coherence 
decides how distinctly visible is the interference pattern.

\begin{figure*}[t!]
\centering
\includegraphics[width=160mm]{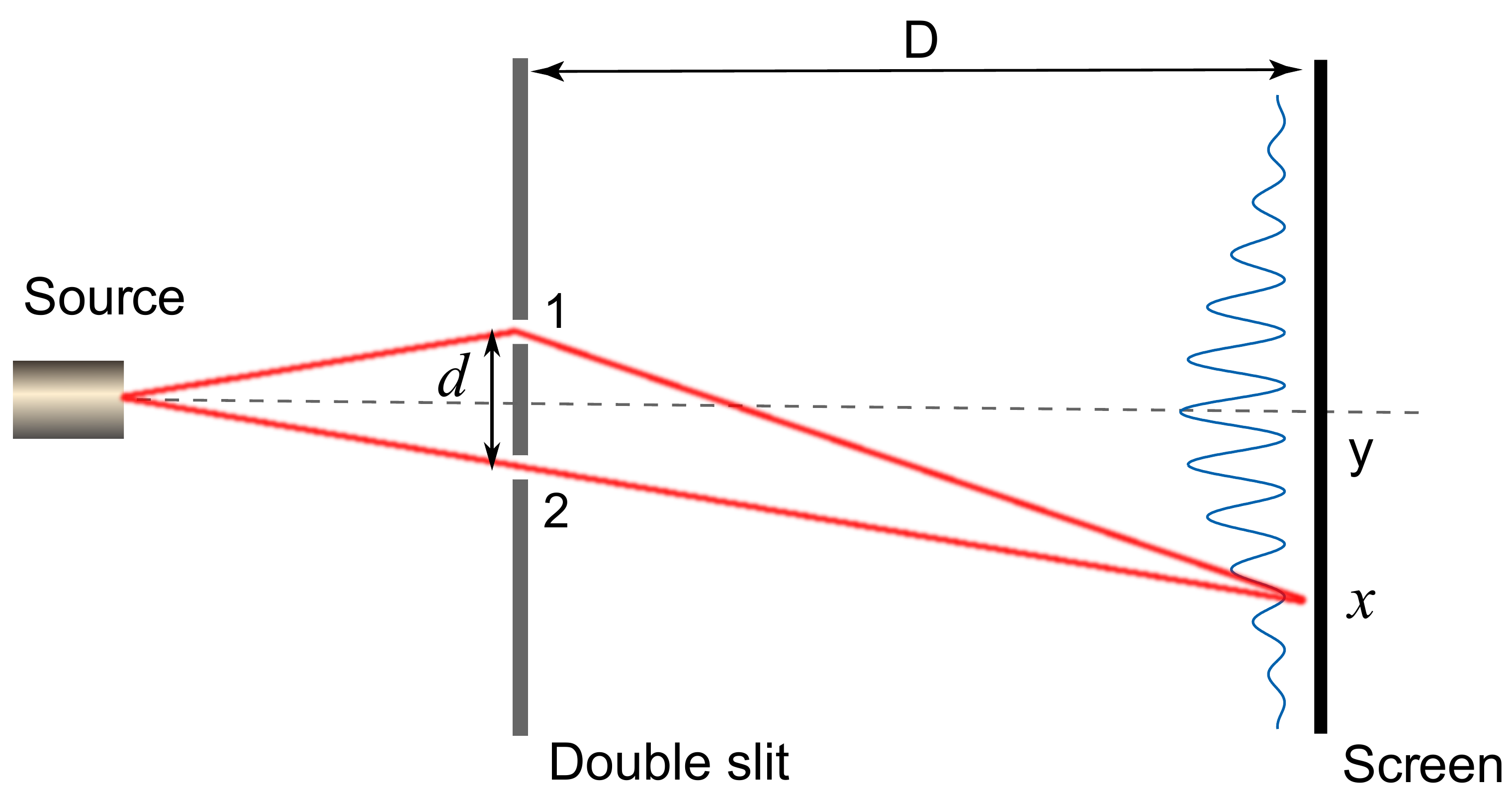}
\caption{Schematic diagram of a two-slit interference experiment.
There are two possible paths a quanton can take, in arriving at the screen.}
\label{twoslit}
\end{figure*}

As the field of classical optics developed, coherence was precisely
defined in terms of a {\em mutual coherence function}, which is
essentially a correlation function \cite{bornw,mandelw}.
If one considers two fields $E_1$ and $E_2$ emanating from two slits, there
is a time difference, say $\tau$, between their arrival at a point on the
screen. The intensity on the screen due to the two slits can be represented as
\begin{eqnarray}
 I &=& I_1 + I_2 + \langle E_1(t)E_2^*(t+\tau)\rangle + \langle E_1^*(t)E_2(t+\tau)\rangle \nonumber\\
   &=& I_1 + I_2 + 2\Re(\langle E_1(t)E_2^*(t+\tau)\rangle) .
\end{eqnarray}
where the angular brackets denote an averaging over $t$, and $I_1$,
$I_2$ represent the respective intensities of the two fields at a
point on the screen.
If $t$ is the time taken for the field to reach the screen from the slits,
it can be represented in terms of the field {\em at the slits}, at an earlier
time:
$E_1(t)=K_1E_1(0),~E_2(t+\tau)=K_2E_2(\tau)$, where $K_1,K_2$ are 
time-indendent propagation functions.
One can define the normalized coherence function, in terms of the field
at the slits, as
\begin{equation}
\gamma_{12}(\tau) = \frac{\langle E_1(0)E_2^*(\tau)\rangle}
{\sqrt{\langle|E_1(0)|^2\rangle \langle|E_2(\tau)|^2\rangle}}
 = \frac{\langle E_1(0)E_2^*(\tau)\rangle}{\sqrt{I_1 I_2}},
\end{equation}

The \emph{visibility of the interference fringes} is conventionally defined as
\cite{bornw}
\begin{equation}
{\mathcal V} = \frac{I_{\textrm{max}} - I_{\textrm{min}}}{ I_{\textrm{max}} + I_{\textrm{min}} } ,
\label{V}
\end{equation}
where the notations have the usual meaning. It can be easily seen that
in this case the visibility turns out to be
$\mathcal{V}=\frac{2\sqrt{I_1 I_2}|\gamma_{12}|}{I_1 + I_2}$, because
$K_1,K_2$ cancel out from the expression. For 
identical width slits which are equally illuminated, the visibility
reduces to
\begin{equation}
\mathcal{V} = |\gamma_{12}|.
\label{vcoh}
\end{equation}
Thus we see that this fringe visibility is a straightforward measure of
coherence of waves coming from the two slits.

Later the field of quantum optics was developed and a quantum theory
of coherence was formulated \cite{glauber,ecg}. However, the quantum theory
of coherence closely followed the earlier classical formulation, except
that the classical fields were replaced by field operators and the 
averages were replaced by quantum mechanical averages \cite{glauberc}. 
The coherence function was still the correlation function of fields.
In analyzing a double-slit interference experiment using quantum optics,
the fringe visibility continued to be related to the mutual coherence
function $\gamma_{12}$ via Eq.~(\ref{vcoh}).

For theoretically analyzing interference experiments done with quantum
particles, e.g. electrons, and experiments like Mach--Zehnder interferometer
where paths traversed are discrete, it is more convenient to use quantum
states and density operators. In such situations, Eq.~(\ref{vcoh}) is
generalized to
\begin{equation}
\mathcal{V} = 2|\rho_{12}|,
\label{vrho}
\end{equation}
where $\rho_{12}$ represents the off-diagonal part of the density matrix of
the quanton, in the basis of states representing quanton in one arm of the
interferometer or the other. The diagonal parts of the density matrix,
$\rho_{11}$ and $\rho_{22}$, represent the probability of the quanton
passing through one or the other arm of the interferometer.

\section{Wave-particle duality}

The fringe visibility given in Eq.~(\ref{V}) formed the basis of all later work
on wave-particle duality. It was taken for granted that the fringe
visibility captures the wave nature, and hence the coherence, of
the interfering quanton, in all interference experiments.
When Bohr proposed his principle of complementarity in 1928 \cite{bohr},
the two-slit interference experiment (see Figure \ref{twoslit}) became a testbed for it. 
For the two-slit experiment, Bohr's principle implies that if one
set up a modified interference experiment in which one gained complete
knowledge about which of the two slits the quanton went through, the
interference would be completely destroyed. 
Acquiring knowledge about which slit the quanton went through, would imply
that the quanton behaved like a particle, by going through one particular
slit. The wave nature was of course represented by the interference pattern.
Wootters and Zurek \cite{wootters} were the first ones to look for a
quantitative statement of Bohr's principle. They studied the effect of 
introducing a path-detecting device in a two-slit interference experiment.
They found that acquiring partial information about which slit the quanton
went through, only partially destroys the interference pattern.
This work was later extended by Englert who derived a wave-particle duality
relation \cite{englert}
\begin{equation}
{\mathcal D}^2 + {\mathcal V}^2 \le 1,
\label{englert}
\end{equation}
where ${\mathcal D}$ is the \emph{path-distinguishability}, a measure of the particle
nature, and ${\mathcal V}$ the visibility of interference, as defined by
Eq.~(\ref{V}). The inequality saturates to an equality if the state of the
quanton and path-detector is pure, unaffected by any external factors like
environment induced decoherence.

The issue of wave-particle duality was looked at, using a different approach,
by Greenberger and Yasin \cite{GY}. In an interference experiment which
is distinctly asymmetric, either because of unequal width of the slits,
or because of the source being unsymmetrically placed with respect to the
two slits, the quanton would be more likely to pass through one of the
slits, than the other. One could make a prediction about which slit the
quanton went through, and would be right more than 50\% of the times.
They argued that the predictability means the quanton is partially behaving
like a particle. They derived the following duality relation for such an
experiment \cite{GY}
\begin{equation}
\mathcal{P}^2 + \mathcal{V}^2 \le 1,
\label{GY}
\end{equation}
where $\mathcal{P}$ is the \emph{path-predictability}, and $\mathcal{V}$ the
visibility of interference, given by Eq.~(\ref{V}).

Subsequently, wave-particle duality for two-path interference has been studied
in various settings
and modifications \cite{einstein,qtwist,coles,vaccaro,awpd,jaeger}, and the same
definition of visibility (\ref{V}) has been used to characterize wave
nature of a quanton.

\section{Multi-path interference}

Jaeger, Shimoni and Vaidman were the first to suggest that one should also
probe complementariy in a $n$-path interference \cite{jaeger}. It is natural to
expect that one should be able to quantitatively formulate Bohr's
complementarity principle for $n$-path interference. Two essential ingredients
needed for such a study would be a definition of distinguishability for
$n$ paths, and probably also a fringe visibility. A lot of effort was made
in this direction \cite{jaeger,durr,bimonte,englertmb,luis,bimonte1,prillwitz},
but a satisfactory $n$-path duality relation remained elusive.

\begin{figure*}[t!]
\centering
\includegraphics[width=170mm]{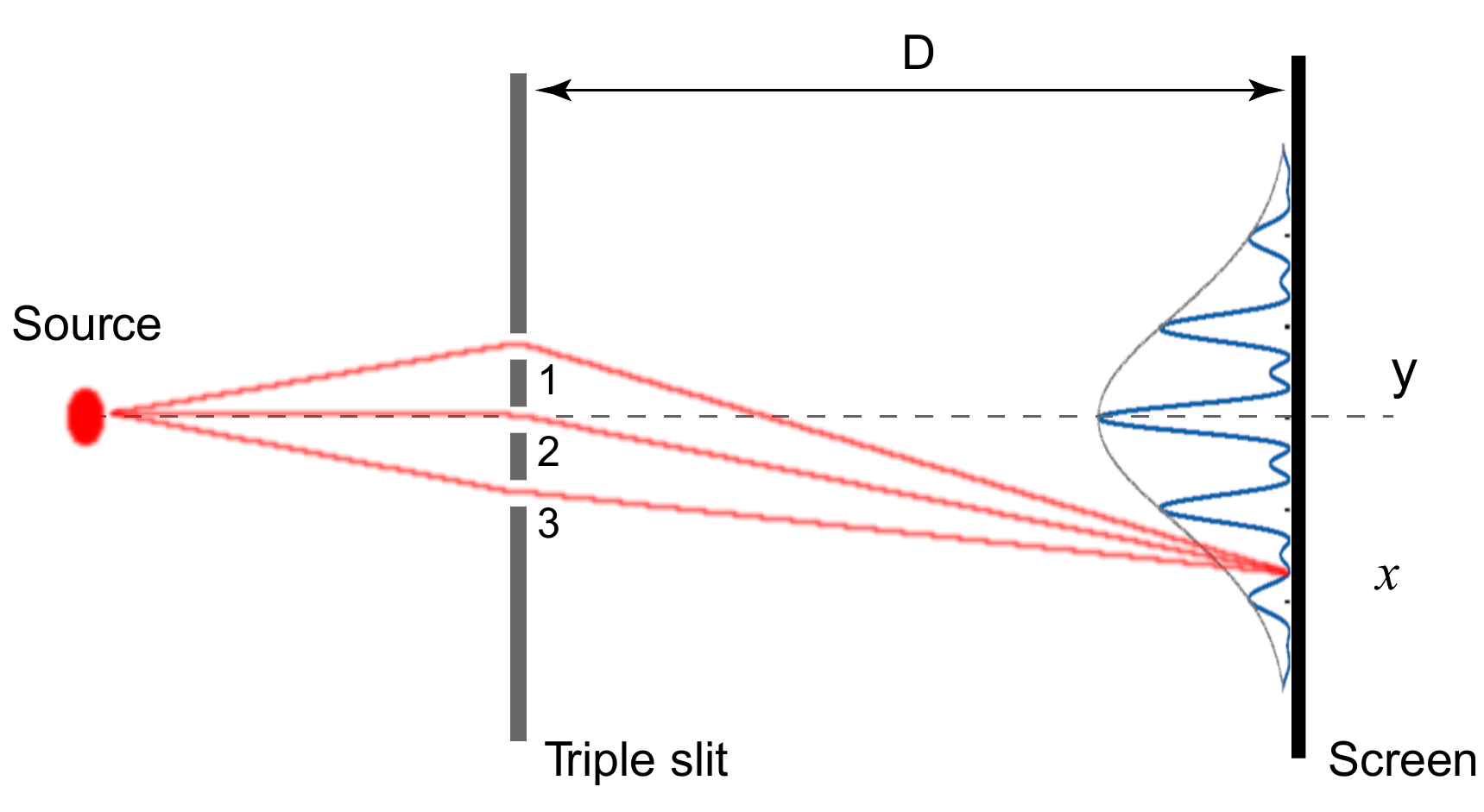}
\caption{Schematic diagram of a three-slit interference experiment.
There are three possible paths a quanton can take, in arriving at the screen.}
\label{threeslit}
\end{figure*}

In 2001, Mei and Weitz carried out multi-beam interference experiments with
atoms, where they scattered photons off selected paths, in order to generate
controlled decoherence \cite{mei}. Surprisingly, they found that there can
be situations where increasing decoherence can actually lead to an increase
in the fringe contrast or visibility, as given by (\ref{V}). The results
seemed to fly in the face of the basic idea of complementarity that any
increase in path knowledge, should lead to a degradation of interference.
However, the authors concluded that the fringe contrast or visibility,
as given by (\ref{V}), is not sufficient to quantify sharpness of
interference.
Based on the results of this experiment, Luis \cite{luis} went to the
extent of claiming that for multi-path quantum interferometers the
visibility of the interference and `which-path' information are not
always complementary observables, and consequently, there are path
measurements that do not destroy the interference.

\subsection{Contrast in multi-beam interference}
\subsubsection{Three-path interference}

In an interesting work, Bimonte and Musto \cite{bimonte1} analyzed the
visibility given by (\ref{V}) for multi-beam interference experiments. They
argued that the traditional notion of visibility is incompatible with any
intuitive idea of complementarity, but for the two-beam case \cite{bimonte1}. 
In the following we rephrase their argument, as it expounds the need for
a new visibility.

Consider a quanton passing through $n$ paths, such that the state corresponding
to the $k$'th path is $|\psi_k\rangle$. This can happen because of a beam-splitter
or because of the quanton passing through a multi-slit. In general,
its density operator may be written as
\begin{equation}
  \rho = \sum_{j=1}^n \sum_{k=1}^n \rho_{jk} |\psi_j\rangle\langle\psi_k|,
\end{equation}
which may in general be mixed. It is quite obvious to see that,
 $\sum_i \rho_{ii}=1$,
where $\rho_{ii}$ represents the fractional population of the $i$'th beam.
Let us assume that the phase in the $j$'th beam
gets shifted by $\theta_j$, such that when the quanton comes out, its state
is
\begin{equation}
  \rho = \sum_{j=1}^n \sum_{k=1}^n \rho_{jk} |\psi_j\rangle\langle\psi_k|
  e^{\imath(\theta_j-\theta_k)}.
\end{equation}
After coming out, the beams are combined and split into new channels,
whose states may be represented by $|\phi_i\rangle$. For simplicity,
we assume that all the original beams have equal overlap with a
particular output channel, say $|\phi_i\rangle$. This amounts to
\begin{equation}
  \langle\phi_i|\psi_1\rangle = \langle\phi_i|\psi_2\rangle =
  \langle\phi_i|\psi_3\rangle = \dots = \langle\phi_i|\psi_n\rangle = \alpha.
\end{equation}
The probability $I$ of finding the quanton in the $i$'th channel is then
given by $I = \langle\phi_i|\rho|\phi_i\rangle$. For the present case,
this probability is given by
\begin{eqnarray}
I &=& |\alpha|^2 \left( \sum_{j=1}^n \rho_{jj} +
\sum_{j \neq k}e^{\imath(\theta_j-\theta_k)} \rho_{jk} \right) \nonumber\\
  &=& |\alpha|^2 \left(1 + 
\sum_{j \neq k}e^{\imath(\theta_j-\theta_k)} \rho_{jk} \right) \nonumber\\
  &=& |\alpha|^2 \left(1 + 
\sum_{j \neq k}|\rho_{jk}| \cos(\theta_j-\theta_k) \right) .
\label{ninterf}
\end{eqnarray}
In order to keep the analysis simple, we assume that all phases in the beams
can be independently varied. That's what allows us to use the absolute value
of $\rho_{jk}$ in the above expression, and absorb all phases associated with
it in $\theta_j, \theta_k$. The probability attains its maximum when
$\theta_j-\theta_k = 2m\pi$ for all $j,k$, where $m$ is an integer. The
condition for the minimum is not as straightforward, and depends in general
on the number of paths $n$. A typical three-slit interference setup and
interference pattern is depicted in Figure \ref{threeslit}.

Next we consider a scenario where the quanton may get entangled with an
ancilla system, which could be an effective environment or possibly a
path-detecting device. If the ancilla system is initially in a state
$|\chi_0\rangle$, the resultant combined state of the quanton and the
ancilla, after their interaction, will be of the form
\begin{equation}
  \rho' = \sum_{j=1}^n \sum_{k=1}^n \rho_{jk} |\psi_j\rangle\langle\psi_k|
  e^{\imath(\theta_j-\theta_k)}\otimes |\chi_j\rangle\langle\chi_k|,
\end{equation}
where $|\chi_i\rangle$ are certain ancilla states, assumed to be normalized,
but not necessarily orthogonal to each other.
Since we are only interested in the behavior of the quanton, we will trace
over the states of the ancilla, to get a reduced density operator
\begin{equation}
  \rho_r' = \sum_{j=1}^n \sum_{k=1}^n \rho_{jk} |\psi_j\rangle\langle\psi_k|
  e^{\imath(\theta_j-\theta_k)} \langle\chi_k|\chi_j\rangle .
\end{equation}
The probability $I'$ of finding the quanton in the $i$'th channel, in
this new case, is given by
\begin{eqnarray}
I'  &=& |\alpha|^2 \left(1 + 
\sum_{j \neq k}|\rho_{jk}|\langle\chi_k|\chi_j\rangle  \cos(\theta_j-\theta_k) \right) .
\label{Ip}
\end{eqnarray}

Now, since the ancilla can get path information, it should always degrade
the interference, except for the trivial case of all $|\chi_i\rangle$ being
identical. Consequently, any meaningfully defined visibility should be
smaller for $I'$ than for $I$. The visibility defined in (\ref{V}) should
then satisfy $\mathcal{V}' \le \mathcal{V}$.
Following Bimonte and Musto, we consider a three-beam interference, where
the density matrix is given by
\begin{eqnarray}
\rho = \frac{1}{3}\begin{pmatrix}
1 & -\lambda & \lambda \\
-\lambda & 1 & -\lambda \\
\lambda & -\lambda & 1 
\end{pmatrix} .
\label{mat}
\end{eqnarray}
For the 3-path interference, maximum intensity occurs when all cosines
are equal to $+1$, and minimum intensity occurs when all cosines are equal
to $-\frac{1}{2}$. Using Eq.~(\ref{mat}), one finds $I_{\textrm{max}}=|\alpha|^2(1+2\lambda)$, and
$I_{\textrm{min}}=|\alpha|^2(1-\lambda)$. Consequently, fringe visibility is
\begin{equation}
\mathcal{V} = \frac{3\lambda}{2+\lambda}.
\end{equation}
One might like to pause here, and compare this relation with the fringe
visibility for two slits, (\ref{vrho}). While the visibility for two slits
was simple, the one for 3-slits contains a denominator too. This is
simply because \mbox{$I_{\textrm{max}}+I_{\textrm{min}}$} sits in the denominator, and is 
independent of $\rho_{jk}$ only for the case of two slits. One can easily
guess that this term will be more complicated once we move on to four or
five slits. 

We next look at the case where the quanton is entangled with the ancilla.
Let us consider the scenario where $\langle\chi_1|\chi_2\rangle=1$ and
$\langle\chi_1|\chi_3\rangle=\langle\chi_2|\chi_3\rangle=0$.
The reduced density matrix, in this case, has the following form
\begin{eqnarray}
\rho_r' = \frac{1}{3}\begin{pmatrix}
1 & -\lambda & 0 \\
-\lambda & 1 & 0 \\
0 & 0 & 1 
\end{pmatrix} .
\label{mate}
\end{eqnarray}
Bimonte and Musto calculated the fringe visibility in this case to yield
\cite{bimonte1}
\begin{equation}
\mathcal{V}' = \frac{4}{3}\lambda,
\end{equation}
and argued that $\mathcal{V}'$ can become larger than $\mathcal{V}$.
One can easily see that this claim is wrong, simply because for 
$\tfrac{3}{4} < \lambda < 1$, $\mathcal{V}'$ becomes larger than 1.
The very definintion of visibility (\ref{V}), guarantees that it cannot be
greater than 1.
Since out of six off-diagonal elements of the density matrix, only two are
non-zero, only one cosine term matters. Maximum intensity will be when
the cosine is $+1$ and minimum, when it is $-1$. This leads to
$I_{\textrm{max}}=|\alpha|^2(1+2\lambda/3)$, and
$I_{\textrm{min}}=|\alpha|^2(1-2\lambda/3)$,
which leads to the correct visibility 
\begin{equation}
\mathcal{V}' = \frac{2}{3}\lambda.
\end{equation}
One can verify that $\mathcal{V}' < \mathcal{V}$ for any value of $\lambda$.

So, although Bimonte and Musto failed to demonstrate the inadequacy of 
Eq.~(\ref{V}) to quantify the wave nature, we will show that is possible to
do that for four-path inteference. However, from the preceding analysis
we are able to show that the expression for visibility gets more complex
as the number of slits increase, and it looks unlikely that one can get
simple duality relations
involving visibility for $n > 2$.

\subsubsection{Four-path interference}

Let us analyze a four-path experiment similar to that studied by Mei and
Wietz \cite{mei}. The intensity is described in general by Eq.~(\ref{Ip}).
Let us a assume a maximally coherent initial state of the quanton:
\begin{equation}
|\psi\rangle = \frac{1}{\sqrt{4}}(|\psi_1\rangle+|\psi_2\rangle+|\psi_3\rangle+|\psi_4\rangle).
\end{equation}
If there is no path-detection or decoherence involved,
$\langle\chi_k|\chi_j\rangle = 1$ for all $j,k$, and the density matrix
of the quanton is given by
\begin{eqnarray}
\rho = \frac{1}{4}\begin{pmatrix}
1 & 1 & 1 & 1 \\
1 & 1 & 1 & 1 \\
1 & 1 & 1 & 1 \\
1 & 1 & 1 & 1 
\end{pmatrix} .
\label{mat4p}
\end{eqnarray}
In order to simulate a typical four-slit interference, let us assume that
all the phases depend on a single parameter $\theta$ such that 
$\theta_k=k\theta, k\in\{1,2,3,4\}$.
Using Eqs.~(\ref{ninterf}) and (\ref{mat4p}), the intensity is now given by
$$I=|\alpha|^2[1+\frac{1}{2}(3\cos \theta 
+ 2\cos 2\theta + \cos 3\theta)].$$
The maximum of the intensity occurs
for $\theta=0$, giving $ I_{\textrm{max}} = 4 |\alpha|^2$. Minimum intensity is
obtained when (say) $\theta=\frac{\pi}{2}$, and is given by
$ I_{\textrm{min}} = 0$. The visibility $\mathcal{V}=1$, as expected.
Now suppose the quanton is entangled with the ancilla, and ancilla states
are such that $|\chi_1\rangle,\chi_2\rangle,\chi_3\rangle$ are all exactly
same, and $|\chi_4\rangle$ is orthogonal to them.
This will lead to $\langle\chi_i|\chi_4\rangle=0, i\in\{1,2,3\}$.
If the ancilla were a
path detector, this situation would imply that the path detector can only
tell if the quanton passed through path 4 or not. It is completely neutral
to all the other three paths.  The reduced density matrix
of the quanton is then given by
\begin{eqnarray}
\rho'_r = \frac{1}{4}\begin{pmatrix}
1 & 1 & 1 & 0 \\
1 & 1 & 1 & 0 \\
1 & 1 & 1 & 0 \\
0 & 0 & 0 & 1 
\end{pmatrix} .
\label{mat4r}
\end{eqnarray}
Using Eqs.~(\ref{Ip}) and (\ref{mat4r}), the intensity is now given by
$$I'=|\alpha|^2[1+\frac{1}{2}(2\cos \theta + \cos 2\theta)].$$
The maximum of the intensity occurs again for $\theta=0$,
and is given by $ I'_{\textrm{max}} = \frac{5}{2} |\alpha|^2$.
Minimum intensity is obtained when $\theta=\frac{2\pi}{3}$, and is given by
$I'_{\textrm{min}} = \frac{1}{4} |\alpha|^2$, which
leads to a reduced visibility $\mathcal{V}' = \frac{9}{11}$.

\begin{figure*}[t!]
\centering
\includegraphics[width=120mm]{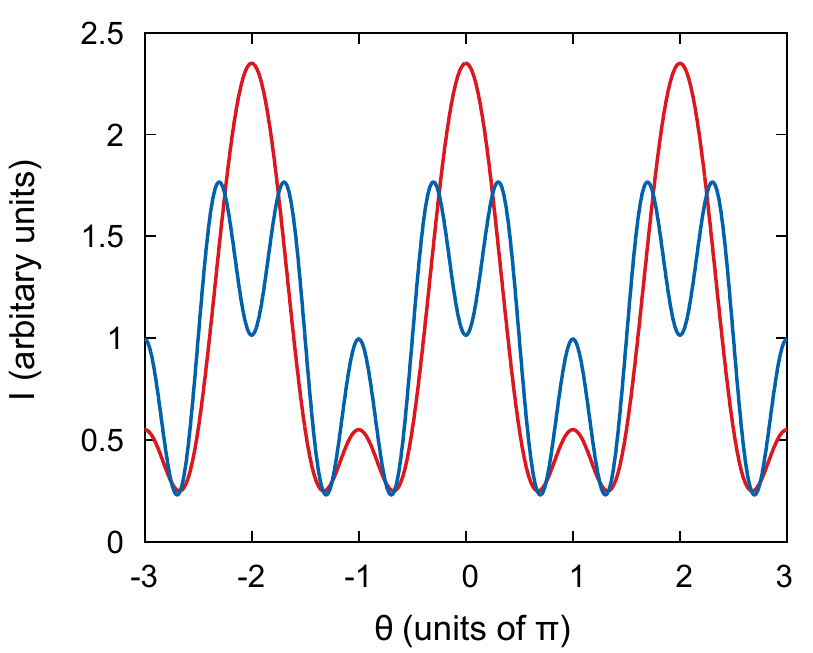}
\caption{A four-path interference pattern, as it appears if the phase in
one path has an additional $\pi$. The blue curve represents interference
when no path-information is obtained. 
The red curve represents interference in the presence of a path detector, 
which can only tell if the quanton passes through path 4. Paradoxically,
the intensity is clearly larger when the path information is available.}
\label{piflip}
\end{figure*}

Now let us consider a new scenario, similar to the one proposed by Mei and
Weitz, where the 4th path is given an additional phase $\pi$, such
that $\theta_4=4\theta+\pi$.
However, the rest of the phases remain as before, $\theta_k=k\theta, k\in\{1,2,3\}$.
The intensity in this scenario is given by
$$I=|\alpha|^2[1+\frac{1}{2}(\cos \theta - \cos 3\theta)].$$
The maximum of
the intensity occurs for $\theta=\frac{\pi}{3}$, and is given by
$I_{\textrm{max}} = \frac{7}{4}|\alpha|^2$.
Minimum intensity is obtained when $\theta=\frac{2\pi}{3}$, and is given by
$I_{\textrm{min}} = \frac{1}{4}|\alpha|^2$.
The visibility then takes the value $\mathcal{V} = \frac{3}{4}$. As
one can see, this is a queer case where even
though the state is pure, and the quanton is equally likely to go through
any path, the visibility is less than 1. Such a thing can never happen
for two-path interference.
Figure \ref{piflip} represent a four-path interference in such a specialized
scenario.

In the presence of the ancilla of the form described in the preceding 
discussion, all the off-diagonal terms involving path 4 are zero, and the 
reduced density matrix is again given by (\ref{mat4r}). Hence the effect of
the addition phase $\pi$ for the fourth path, also disappears here.
The intensity now is given by
$$I'=|\alpha|^2[1+\frac{1}{2}(2\cos \theta + \cos 2\theta)].$$
The maximum of
intensity is obtained when $\theta=0$, leading to
$I'_{\textrm{max}} = \frac{5}{2}|\alpha|^2$. Minimum intensity is obtained when
$\theta=\frac{2\pi}{3}$, leading to $I'_{\textrm{min}} = \frac{1}{4}|\alpha|^2$.
Thus the visibility of interference,
in the presence of the ancilla, is $\mathcal{V}'=\frac{9}{11}$.
Compare this with the visibility
without the ancilla, $\mathcal{V} = \frac{3}{4}=\frac{9}{12}$, and we get a very
counter-intuitive  result, $\mathcal{V}' > \mathcal{V}$. Getting selective
path information about the
quanton, increases the fringe visibility. Bohr's complementarity principle
implies that getting any path information about the particle, should always
decrease the wave nature of the quanton. Assuming that Bohr's principle should
always hold true, we conclude that visibiliy $\mathcal{V}$, as given by
(\ref{V}), is not a good measure of the wave nature of a quanton in multi-path
interference.

\subsection{A duality relation for 3-slit interference}

In a radically different approach to quantify path-knowledge, a new
path-distinguishability
was introduced for three-slit interference \cite{3slit}, based on
\emph{unambiguous quantum state discrimination} \cite{uqsd1,uqsd2,uqsd3,uqsd4}. The new
path distinguishability is denoted by $\mathcal{D}_Q$, and duality relation
for three-slit interference, that was derived, is
\begin{equation}
\mathcal{D_Q} + {2\mathcal{V}\over 3- \mathcal{V}} \le 1 ,
\end{equation}
where the visibility $\mathcal{V}$ is same as (\ref{V}). This duality 
relation correctly generalizes Bohr's principle of complementarity to
three-slit interference. However, the elegance seen in the duality
relation for two-slit interference, (\ref{englert}), is missing from its form.
One might suspect that it might suffer from the problems pointed out by
Mei and Weitz \cite{mei}, but that has not been demonstrated.

\subsection{Coherence: a new measure of wave nature}

As argued earlier, in quantum optics, coherence was formulated in terms
of correlation function of field operators. This approach works quite
well for quantum optics, however a general definition of coherence,
grounded in quantum theory, was missing. 
A new measure of coherence was introduced by Baumgratz, Cramer and Plenio in a
seminal paper \cite{coherence}. It is just the $\ell_1$ norm of the 
off-diagonal elements of the density matrix, in a particular basis.
In the context of multi-path interference, the
basis states can be the states of the quanton corresponding to its passing 
through various paths. Based on Baumgratz, Cramer and Plenio's measure, one
can define a normalized coherence as \cite{cd}
\begin{equation}
{\mathcal C} \equiv {1\over n-1}\sum_{j\neq k} |\rho_{jk}| ,
\label{C}
\end{equation}
where $\rho_{jk}$ are the matrix elements of the density operator of the
quanton in a particular basis, and $n$ is the dimensionality of the
Hilbert space. In the context of multi-path interference, $n$ would be
the number of slits and the basis states would be the states corresponding
to various paths the quanton can take. The value of $\mathcal{C}$ is
bounded by $0 \le \mathcal{C} \le 1$.

It has been argued that this measure of quantum coherence, can be a good
quantifier of wave nature of the quanton. We test out $\mathcal{C}$ for
the previous case of 4-path interference with one path having an additional
phase $\pi$, where $\mathcal{V}$ gave a counter-intuitive result.
From the definition of $\mathcal{C}$ one can see that it does not
depend on the phases at all. Coherence for this case is given by
\begin{equation}
{\mathcal C} = {1\over 4-1}\sum_{k(\neq j)=1}^4\sum_{j=1}^4 |\rho_{jk}| = 1 .
\end{equation}
So, coherence turns out to have its maximum possible value $\mathcal{C}=1$
for this pure quanton state, as it should be. Contrast this with
$\mathcal{V}=\frac{3}{4}$ for the same case.
Next let us calculate the coherence for the case where the ancilla is
also present, the case represented by (\ref{mat4r}). It is straightforward
to calculate
\begin{equation}
\mathcal{C}' = {1\over 4-1}\sum_{k(\neq j)=1}^4\sum_{j=1}^4 |{\rho'_{r}}_{jk}|
= \frac{1}{2} .
\end{equation}
So we get $\mathcal{C}' < \mathcal{C}$, in full agreement with Bohr's principle
of complementarity. Coherence turns out to be a better quantifier of wave
nature, as compared to conventional visibility, in this respect.

Coherence $\mathcal{C}$ should then be the right measure to be used in
formulating wave-particle duality relations for multi-path interference.
The new path-distinguishability defined for three-slit interference
can be generalized to $n$-slits \cite{3slit}. This distinguishability,
when combined with coherence $\mathcal{C}$, yields an elegant wave-particle
duality relation \cite{cd}
\begin{equation}
\mathcal{D}_Q + \mathcal{C} \le 1,
\label{cd}
\end{equation}
for $n$-path interference. The form of this duality relation is different from
Eq.~(\ref{englert}). However, one can also formulate a $n$-path duality relation
which is of the same form as Eq.~(\ref{englert}), if one uses a different
definition of path-distinguishability (see \cite{nslit})
\begin{equation}
\mathcal{D}^2 + \mathcal{C}^2 \le 1.
\label{nslit}
\end{equation}

Very recently, a duality relation between path-predictability (not
path-distinguishability) and coherence has also been formulated for
$n$-path interference (see \cite{predict})
\begin{equation}
\mathcal{P}^2 + \mathcal{C}^2 \le 1,
\label{predict}
\end{equation}
where $\mathcal{P}$ is a path-predictability. This relation generalizes the duality
relation of Greenberger and Yasin (\ref{GY}) to $n$-path interference.
All three of these duality relations saturate for all pure states. 
Apart from the above works, there have been other investigations into 
wave-particle duality for multi-path interference \cite{coles1,bagan}.

Coherence has proved to be a very versatile tool and has been
used in studying a variety of situations where it arises due to the lack of
distinguishability which is different from path distinguishability
in multi-path interference. For example, recently it has been shown that
in double parametric pumping of a superconducting microwave cavity,
coherence between photons in separate frequency modes can arise because of
the absence of which-way information in the frequency space \cite{paraoanu1}.
Similarly, in a double spontaneous down-conversion processes, it has been
shown that coherence in photons in the signal is induced due to lack of
which-path information for the photons emitted in the idler \cite{paraoanu2}.
Coherence has also found applications in quantum metrology \cite{qmetro}.
Coherence arising out of indistinguishability of identical particles
has been shown to be useful in quantum metrology \cite{castellini}.

\subsection{Coherence as a new visibility}

\begin{figure*}[t!]
\centering
\includegraphics[width=125mm]{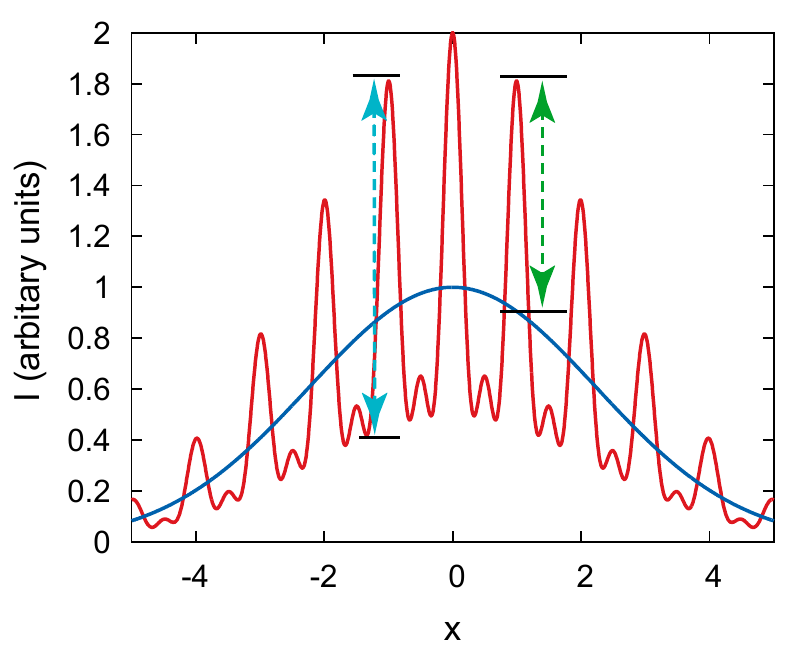}
\caption{A typical unsharp three-slit interference pattern is shown here.
When the interference is lost, the broad Gaussian profile (shown in the figure) is what remains.
Traditional visibility is based on the intensity difference depicted
by the dashed line to the left of the central maximum. A new visibility
can be defined by the intensity difference depicted by the dashed line
to the right of the central maximum.}
\label{interf}
\end{figure*}

With the three duality relations described in the preceding subsection,
it appears that the problem of generalizing the quantitative statement of
Bohr's complementarity, to \mbox{$n$-path}~interference, is solved. However, one
might still ask how this coherence $\mathcal{C}$ is connected to interference.
This question has been the subject of some recent investigations \cite{biswas,tania}.
To address this question we again consider a quanton going through $n$
paths. Equation (\ref{ninterf}) gives us the probability for a quanton
to be found in a particular output channel, which looks like the following:
\begin{eqnarray}
I  &=& |\alpha|^2 \left(1 + 
\sum_{j \neq k}|\rho_{jk}| \cos(\theta_j-\theta_k) \right) .
\label{I}
\end{eqnarray}
The second term in the large brackets is the one which signifies interference.
In addition to the interference between various paths, there is also a
probability associated with the quanton passing through an individual path.
The first term just represents the sum of these probabilities corresponding
to each path. A typical unsharp interference pattern is shown in
Figure \ref{interf}. If the interference is absent, one would only
see a broad Gaussian distribution of intensity, represented by the
blue curve in Figure \ref{interf}.

Conventional visibility, given by Eq.~(\ref{V}), is calculated using the
intensity difference between a maximum and a nearby minimum, represented
by the dashed line on the left of the central maximum in Figure \ref{interf}. 
Now suppose we define a {\em new visibility} $\mathcal{V}_{C}$ by taking
the difference
between the intensity at a primary maximum and the intensity at the
same position if there were no
interference. This is represented by the dashed line to the right of the
central maximum in Figure \ref{interf}. We would like to scale this
difference with the intensity at the same position if there were no
interference. In addition, we would like to scale it with $(n-1)$, the
reason for which will become clear later.
So, our new visibility will look like the following:
\begin{equation}
\mathcal{V}_{C} = \frac{1}{n-1} \frac{I_{\textrm{max}} - I_{\textrm{inc}}}{ I_{\textrm{inc}} } ,
\label{Vn}
\end{equation}
where $I_{\textrm{inc}}$ represents the intensity at the position of a primary maximum 
if the source is made incoherent, and {\em the interference is destroyed}.
How would one obtain $I_{\textrm{inc}}$?
One example may be by introducing a phase randomizer in the path of light
before it enters the multi-slit, something that is used in some modern
optics experiments. From our example (\ref{I}), these intensities can be
calculated. As mentioned before, $I_{\textrm{max}}$ corresponds to the intensity
when all the cosines are equal to +1. If some phases of the paths are 
fixed in such a manner that making all cosines equal to +1 is not possible,
this analysis cannot be used. The following can then be easily inferred.
\begin{eqnarray}
I_{\textrm{max}}  &=& |\alpha|^2 \left(1 + \sum_{j \neq k}|\rho_{jk}| \right) \nonumber\\
I_{\textrm{inc}}  &=& |\alpha|^2 .
\label{Is}
\end{eqnarray}
Using (\ref{Is}) the new visibility can be written as
\begin{equation}
\mathcal{V}_{C} = \frac{1}{n-1}
 \frac{ |\alpha|^2 \left(1 + \sum_{j \neq k}|\rho_{jk}| \right) - |\alpha|^2}
{|\alpha|^2}
= \frac{1}{n-1} \sum_{j \neq k}|\rho_{jk}|.
\label{Vnew}
\end{equation}
Comparing with Eq.~(\ref{C}) we see that this is just the coherence of the
quanton, $\mathcal{C}$. So, the new way of measuring visibility yields the
value of coherence. In other words, in multi-path intereference, coherence is
a measurable quantity and can be inferred by measuring the intensities of the
interference, although by a more involved method. What is presented here,
is an idealized version of interference. In reality the incoming state could
be a wave-packet, and the slits could be of finite width. In addition,
one may wonder if the position, at which the intensities are measured,
makes a difference to the visibility. All these issues have been addressed
in a more detailed analysis \cite{tania}.

For two-slit interference we can compare $\mathcal{V}_{C}$ with $\mathcal{V}$.
For $n=2$, the new visibilty is given by
\begin{equation}
\mathcal{V}_{C} = \frac{1}{n-1} \sum_{j \neq k}|\rho_{jk}| = 2|\rho_{12}|,
\label{Vc2}
\end{equation}
which is exactly the same as the traditional visibility (\ref{vrho}).
So, for two-slit interference, $\mathcal{V}_{C}$ has the same value as
$\mathcal{V}$. This is something that nobody could have guessed, that two
different ways of measuring visibility will yield the same answer.

Getting an experimental handle on coherence was a difficult task, mainly
because it is defined only in terms of the density matrix, in a particular
basis. In a multislit experiment for example, the basis is not well defined,
not to speak of the elements of the density matrix. 
The prospect of measuring coherence from the interference pattern has
opened up new possibilities \cite{anu,misba,gong}.

Coherence can be experimentally measured by the method described in the
preceding discussion in most scenarios. However, in an experiment like
that of Mei and Weitz, the phases of the paths are arranged in such a
manner that getting path information changes the very character of
the interference pattern, instead of just degrading it. In such a 
situation the procedure to measure coherence from the interference
pattern in Ref. \cite{tania}, fails. As one can see from Figure \ref{piflip},
the interference does become sharper, as path information is extracted.
However, we have seen from the preceding discussion that coherence 
$\mathcal{C}$ does decrease with increasing path information, even though
the interference becomes sharper. So, it might look like there is no
hope of getting $\mathcal{C}$ from the interference in such a situation.
However, very recently it has been demonstrated that in multi-path
interference, coherence can be measured from interference in a rather
unconventional way \cite{cdsum}. The idea is the following. Instead of
measuring the visibility of a multi-path interference directly, one
blocks $n-2$ paths, and lets only two paths $i,j$ be open. Then what
one gets is just a two-path interference, whose visibility is given
by (\ref{vrho}) as
\begin{equation}
\mathcal{V}_{ij} = \frac{2|\rho_{ij}|}{\rho_{ii}+\rho_{jj}}.
\label{Vij}
\end{equation}
One can repeat this procedure by opening another pair of paths, and 
measure $\mathcal{V}_{ij}$ for that. If all paths are equally probable,
$\rho_{ii}=\rho_{jj}=\frac{1}{n}$. The average visibility of all the
$n(n-1)/2$ pairs of paths is given by
\begin{equation}
\mathcal{V} = \tfrac{2}{n(n-1)}\sum_{\text{pairs}}\mathcal{V}_{ij}
= \tfrac{2}{n(n-1)}\sum_{\text{pairs}}\frac{2|\rho_{ij}|}{\rho_{ii}+\rho_{jj}}
= \tfrac{1}{n-1}\sum_{i\ne j} |\rho_{ij}|,
\label{Vsum}
\end{equation}
which is exactly the coherence $\mathcal{C}$ defined by (\ref{C}).
So, coherence can be measured as the average two-path visibility over
all path pairs, by selectively opening only one pair of paths at a time.
If all paths are not equally probable, coherence can be measured in a
slightly different way \cite{cdsum}
\begin{equation}
\mathcal{C} = \tfrac{1}{n-1}\sum_{\text{pairs}} (\rho_{ii}+\rho_{jj})\mathcal{V}_{ij},
\end{equation}
where $\rho_{ii}$ can be interpreted as the measured relative intensity in the $i$'th path.
This method of measuring $\mathcal{C}$ works in all kinds of multi-path
interference experiments, even in an experiment like that of Mei and Weitz \cite{mei}.

One might wonder if, what has been looked at in the preceding analysis, is
enough to accord the status of a new visibility to coherence, or may
there be some more conditions still desired. A very thorough analysis of the issue
had been done by D\"{u}rr while looking for a new visibility
and new predictability for multi-path interference. He suggested that
any newly defined visibility should satisfy the following
criteria \cite{durr}\\[6pt]
(1) It should be possible to give a definition of visibility that is
based only on the interference pattern, without explicitly
referring to the matrix elements of $\rho$ .\\[6pt]
(2) It should vary continuously as a function of the matrix
elements of $\rho$ .\\[6pt]
(3) If the system shows no interference, visibility should reach its global
minimum.\\[6pt]
(4) If $\rho$ represents a pure state (i.e., $\rho^2 = \rho$) and all $n$~beams
are equally populated (i.e., all $\rho_{jj}=\frac{1}{n}$), visibility should
reach its global maximum.\\[6pt]
(5) Visibility considered as a function in the parameter space
($\rho_{11}, \rho_{12},\dots , \rho_{nn}$) should have only global extrema,
no local ones.\\[6pt]
(6) Visibility should be independent of our choice of the coordinate system.\\[6pt]
Coherence, as defined by Eq.~(\ref{C}), satisfies all of D\"{u}rr's criteria.
Thus, one can confidently accord it the status of visibility for multi-path
interference.

\section{Discussion}

When things started out with interference in classical waves, coherence was
the quantity that turned out to be related to the visibility in
two-path interference. As quantum optics developed, Glauber's coherence
function \cite{glauberc} turned out to be related to visibility of
interference of light on the quantum scale. Later, when interference of
particles started being analyzed, using quantum states, coherence was
dropped and the visibility took its place. The reason for this was, probably,
a missing general enough definition of coherence in the quantum domain.
This was one of the stumbling blocks which stalled the generalization of
Bohr complementarity, from two-path to multi-path interference.
The other stumbling block was that the way in which path-distinguishability was
defined by Englert \cite{englert}, provided no natural way for generalization
to the multi-path scenario. A new definition of path-distinguishability,
based on unambiguous quantum state discrimination, provided a natural generalization to multi-path
distinguishability \cite{3slit}.

A new definition of coherence by Baumgratz, Cramer and Plenio \cite{coherence}
removed the stumbling block in quantifying the wave nature in multi-path
interference experiments. It led to the formulation of universal wave-particle
duality relations for multi-path interference \cite{cd,nslit,predict}.
Now that coherence given by Eq.~(\ref{C}) has proved to be a good visibility of
interference, coherence can again be accorded the position of the quantifying
measure of the wave nature of quantons.

Lastly, in certain specialized scenarios in multi-path interference, where
the phases in the paths cannot be varied independently, like the one 
represented by Figure \ref{piflip}, the interference pattern itself may not be
able to represent the wave nature of the quanton properly. 
We recall the conclusion of Luis \cite{luis}, namely, that there are path
measurements that do not destroy intreference. Contrary to that, we would
like to stress that any path measurement will necessarily degrade the
{\em coherence} of the quanton.
Coherence remains a good measure of wave nature even in such scenarios.
Not only that, it can always be measured from interference, although in a 
slightly nontrivial way \cite{cdsum}.

\section*{Acknowledgment}
The author acknowledges useful discussions with Anu Venugopalan and
Sandeep Mishra on controlled decoherence in multi-path interference,
and computational support from Imtiyaz Ahmad Bhat.

\bibliography{references}
\bibliographystyle{quanta}

\end{document}